\documentclass[10pt,letterpaper,twocolumn,aps,showpacs,preprintnumbers,amsmath,amssymb,nofootinbib,prl]{revtex4-1}
\usepackage[latin9]{inputenc}
\usepackage{color}
\usepackage{amsmath}
\usepackage{amssymb}
\usepackage{graphicx}
\usepackage[unicode=true,pdfusetitle,
 bookmarks=true,bookmarksnumbered=false,bookmarksopen=false,
 breaklinks=false,pdfborder={0 0 0},pdfborderstyle={},backref=false,colorlinks=true]
 {hyperref}
\hypersetup{
 pdfborderstyle={},colorlinks,linkcolor=red,citecolor=blue}

\makeatletter

\usepackage{setspace}
\usepackage{dcolumn}
\usepackage{mathrsfs}
\usepackage{bm}

\makeatother

\begin{document}
\title{Non-Hermitian magnon-photon interference in an atomic ensemble}
\author{Rong Wen,$^{1}$ Chang-Ling Zou,$^{5}$ Xinyu Zhu,$^{1}$ Peng Chen,$^{1}$
Z. Y. Ou,$^{4}$ J. F. Chen,$^{1,3,\dagger}$ and Weiping Zhang$^{2,3,\dagger}$}
\affiliation{$^{1}$Quantum Institute for Light and Atoms, Department of Physics,
East China Normal University, Shanghai 200241, China~~\\
 $^{2}$Department of Physics and Astronomy, Shanghai Jiao Tong University,
Shanghai 200240, China~~\\
 $^{3}$Collaborative Innovation Center of Extreme Optics, Shanxi
University, Taiyuan, Shanxi 030006, China~~\\
 $^{4}$Department of Physics, Indiana University-Purdue University
Indianapolis, 402 North Blackford Street, Indianapolis, Indiana 46202,
USA~~\\
 $^{5}$Key Laboratory of Quantum Information, University of Science
and Technology of China, Hefei, 230026, China~~\\
 $^{\dagger}$e-mail: jfchen@phy.ecnu.edu.cn; wpzhang@phy.ecnu.edu.cn}
\date{\today}
\begin{abstract}
The beam-splitter (BS) is one of the most common and important components
in modern optics, and lossless BS which features unitary transformation
induces Hermitian evolution of light. However, the practical BS
based on the conversion between different degree of freedoms are
naturally non-Hermitian, as a result of essentially open quantum dynamics.
In this work, we experimentally demonstrate a non-Hermitian BS
for the interference between traveling photonic and localized magnonic
modes. The non-Hermitian magnon-photon BS is achieved by the coherent
and incoherent interaction mediated by the excited levels of atoms,
which is reconfigurable by adjusting the detuning
of excitation. Unconventional correlated interference pattern is observed at the
photon and magnon output ports. Our work is potential for extending to single-quantum level to realize interference between a single photon and magnon, which provides an efficient and
simple platform for future tests of non-Hermitian quantum physics.
\end{abstract}
\maketitle
An optical linear beam-splitter mixes the photons in different paths
to interfere, and it is an essential device in various optical applications,
such as the gravitational wave detection \cite{LIGO} and optical coherence
tomography \cite{Lvovsky}. A BS can also manipulate photonic
quantum state encoded in different modes, which is the most basic
device for building quantum communication network \cite{Hariharan1991,Zukowski1993,Walther2004,Lenef1997}.
Such a concept in photonics have also been generalized to other excitations,
including the phonon \cite{Toyoda2015}, plasmon \cite{Fakonas2014}, magnon \cite{AVChumak},
and matter waves \cite{Lopes2015}. However, almost all of previous studies
assume the Hermitian interactions in a closed quantum system. Considerable
works demonstrate that it is possible to change the phase factor of
the reflection and the transmission coefficients when considering
losses in a BS \cite{Barnett1998}, and
the quantum operations on photons would be affected by such non-Hermitian
system \cite{Zou2017}. Until recently, the unconventional coalescence
have been demonstrated for interfering single plasmons in a non-unitary
BS system \cite{Vest2017}.

While the non-Hermitian interference is of great importance conceptually
and fundamentally, realizing a full controllable non-Hermitian BS
is still a challenge. It is anticipated that any conversion process
between excitations could be treated as a BS \cite{Ou2008,Reim}, therefore a hybrid atom-light interface
provides an appealing and versatile platform to study the photon-matter
interference \cite{Qiu2016,Campbell2012}. Driven by an external optical field,
magnon in an atomic ensemble, also known as collective
atomic spin-wave excitation \cite{Fleischhauer2000}, could be converted
to photon. Such an magnon-photon beam-splitter (MPBS) have been
extensively studied for quantum storage applications \cite{Lukinstorage,Matsukevich}
and also are significant media for entangling distant quantum node
through light \cite{DLCZ}.

\begin{figure}
\includegraphics[width=1\columnwidth]{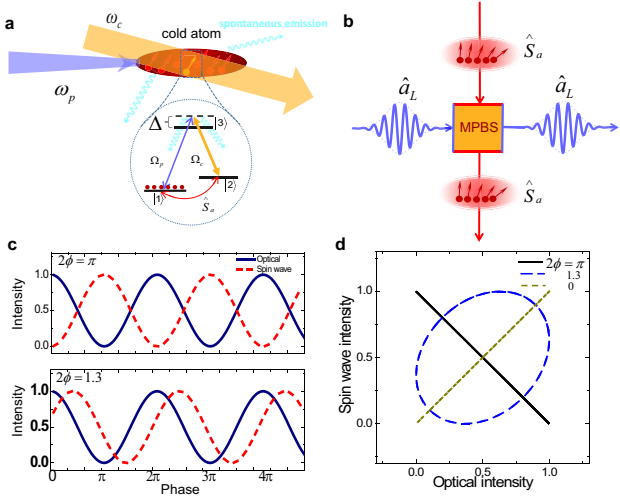} \caption{\textbf{Schematics of the non-Hermitian MPBS.} (a)
photon-magnon conversion in a three-level atom, where a control beam
can stimulate the coherent conversion between the input probe photons
and collective atomic excitation between the two metastable states
$|1\rangle$ and $|2\rangle$ (magnon), while the spontaneous emission
induces an incoherent conversion between photon and magnon. The inset:
three-level $\Lambda$ atomic energy level. Ground state $|1\rangle$
($|2\rangle$): 5$S_{1/2},F=2$ ($F=3$); Excited state $|3\rangle$:
$5P_{1/2},F=3$. The probe beam and control beam are two-photon resonant.(b)
The cold atom ensemble serves as a beam-splitter, where the input
optical wave and spin wave interfere. (c) The sinusoidal fringes of
interference between the optical wave and spin wave. (d) The phase
diagram for the photon and magnon output, with the black solid and
green dot lines presenting the phase difference of fringe $2\phi=\pi$
and $0$, respectively. The ellipse represents an arbitrary non-Hermitian
beam-splitter. }
\label{Fig1}
\end{figure}
In this Letter, we realized a non-Hermitian MPBS mixing photons and
magnons in cold atomic ensembles. By an ensemble of three-level atoms
with near-resonant control laser, reconfigurable coherent and incoherent
conversion between traveling photonic and localized magnonic modes
is demonstrated. By a temporal Mach---Zehnder inteferometer, the
interference based on such MPBS have been demonstrated, and the interesting
unconventional phenomena of non-Hermitian interference are revealed.
Our study provides an versatile experimental platform to study the
non-Hermtian physics and parity-time symmetry at single quanta level
\cite{Jones-Smith2010,El-Ganainy2018}. By exploiting the multiple
level atoms and optical control, our results can be generalized to
multiple magnon mode and photons with different color, and even bilinear-form
of interactions to study the interplay between gain and loss\cite{Hudelist}.

\noindent
\begin{figure*}
\includegraphics[width=2\columnwidth]{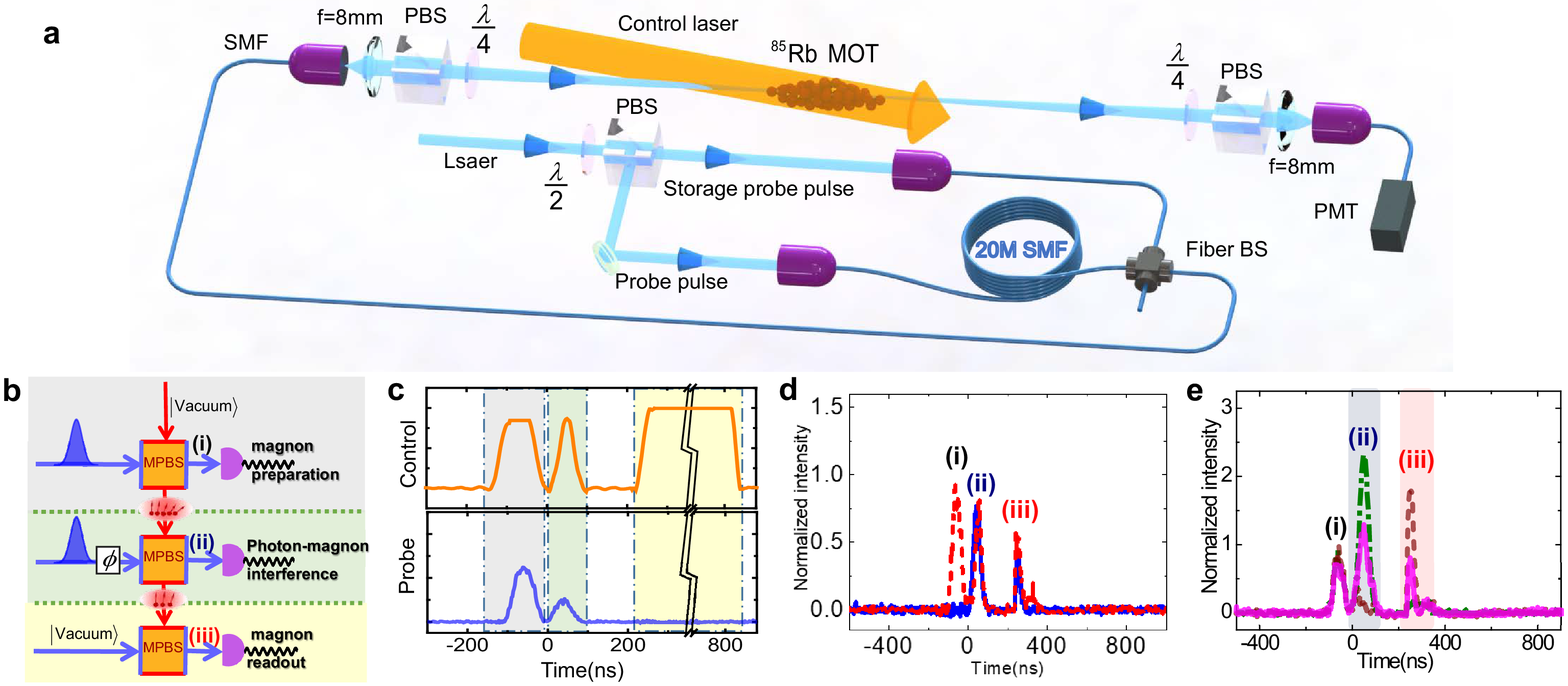} \caption{\textbf{Experimental setup.} (a) The weak probe pulse splitting into
two pulse are launched into an $^{85}$Rb MOT (with 20-meter-fiber
inserted) along the longitudinal axis of the atomic cloud, the angle
between the control beam and the atom long axis direction is $2.5^{\circ}$.
the output for the photon is directly collected by a photomultiplier
tube (PMT). (b) Schematic diagram of the Photon-magnon interference.
(c) The timing sequence of the control light and the probe light represented
by the orange and the blue line. (d) Single port input to MPBS (either
photon or magnon be vacuum state) with balanced splitting ratio represented
by the blue solid line and the red dash line respectively. (e)photon-magnon
interference at different interferometric phase (show 3 results in
the figure) when both the optical and spin wave exist. }
\label{Fig2}
\end{figure*}

As sketched in Fig.$\,$\ref{Fig1}(a), a cigar-shaped laser cooled
$^{85}\mathrm{Rb}$ atoms cloud is prepared in a two-dimensional magneto-optical
trap (MOT) \cite{Du2013,Zhang2015}, and the non-Hermitian MPBS is
realized by external control laser stimulating conversion between the
signal photon and magnon. The underlying quantum processes is explained
by the energy level diagram (inset of Fig.$\,$\ref{Fig1}(a)): a
strong control laser ($\Omega_{c}$) induces coherent conversion between
$\left|2\right\rangle $ and $\left|3\right\rangle $, and also initialize
the atom ensemble in the ground state $\left|1\right\rangle $; signal
photons ($\Omega_{p}$) can be converted to magnon, i.e. the excitation
on level $\left|2\right\rangle $, through the intermediate excite
state. Due to the spontaneous emission of atomic excited state, there
are two incoherent processes associate with the MPBS operation: (i)
direct exciting atoms to the excited state, the signal photon be absorbed
and scattered to free space; (ii) the population on $\left|2\right\rangle $
be further excited to excited state by the control laser, and decay
to $\left|1\right\rangle $ by emitting a photon into free space.
Although the two decoherence processes would lead to the loss of photon
and magnon separately, they are not independent because both processes
rely on the same excited state and free-space optical modes. Therefore,
the two decoherence processes would interfere on the shared decay
channel, allows a indirect magnon-photon conversion process mediated
by the decay channels. As a result, the spontaneous emission of excited
state contributes an effective incoherent interaction between magnon
and photon, which laid as the foundation of the non-Hermitian MPBS
studied in this work.

Fig. \ref{Fig1}(b) conceptually illustrates the MPBS: a traveling
signal photon pulse mode ($\hat{a}_{L}$) and a localized magnon mode
($\hat{S}_{a}$) are mixed via the MPBS. By preparing input photon
and magnons with different initial phase, and interfere them on the
MPBS, the intensity of output photon and magnon would show sinusoidal
fringes. In Fig.$\,$\ref{Fig1}(c), the results for a ideal Hermitian
MPBS is plotted, where fringes for photon and magnon show complementary
oscillations due to the energy conservation. For a general MPBS, we
can introduce a phase factor $\phi$ for the reflection coefficients ($\left|r\right|e^{i\phi},\,\left|r^{\prime}\right|e^{i\phi}$),
with transmission coefficient be real. Therefore, $2\phi=\pi$
for unitary MPBS, and be other values for non-Hermtian
MPBS \cite{Zeilinger1981}. By plotting the Lissajous curve
of the outputs from two-port (Fig.$\,$\ref{Fig1}(d)), it is found
that the non-Hermitian MPBS can change the outputs from anti-correlated
$\left(2\phi=\pi\right)$ to correlated $\left(\phi=0\right)$.

The photon-magnon interference in non-Hermitian MPBS is performed by
the experimental setup shown in Fig.~\ref{Fig2}(a). The schematics
of the temporal inteferometric setup is shown in Fig.~\ref{Fig2}(b).
It is worth noting that the magnon port is virtual because we do not
directly access the magnon, therefore our MPBS is used for interference
in time-domain instead of spatial, and the MPBS is implemented three
times for magnon state preparation, interference and magnon state
readout, respectively. To prepare the magnon, a weak probe pulse is
sent to the atomic cloud and converted to the collective ground state
spin excitation, which is equivalent to a quantum storage process
\cite{Hsiao2018}. To realize the interference by the MPBS, another
probe pulse is sent into the atom after the magnon preparation. The
MPBS is enabled by the control laser, and the output for the photon
is directly collected by a photomultiplier tube (PMT). The output
of the magnons are indirectly measured by converted them to photons
using another control laser pulse, and such a magnon readout process
that depletes the magnon excitation in atomic cloud also serves as
an initialization process for next experiment cycle.

Figs~\ref{Fig2}(d)-(e) show the outputs of the MPBS for the case
of single photon detuning $30\,\mathrm{MHz}$ with an optical depth
(OD) of 30 as an example. In the following experiments, we adjusted
the coupling pulse intensities for the magnon preparation and interference
steps to match the amplitudes of output photon and magnon \cite{Zhu2018}.
Figure~\ref{Fig2}(d) shows the results for single port input to
MPBS (either photon or magnon be vacuum state), and Fig.~\ref{Fig2}(e)
demonstrates the interference results captured at different relative
phase of between the two arms of inteferometer. The output intensities
of photon ($n_{a}=\langle\hat{a}_{L}^{\dag}\hat{a}_{L}\rangle$) and
magnon ($n_{s}=\langle\hat{S}_{a}^{\dag}\hat{S}_{a}\rangle$) can
hence be obtained by integrating the areas of the output pulses (grey
region and the rosiness region) in the Fig.~\ref{Fig2}(e).

\begin{figure}
\includegraphics[width=1\columnwidth]{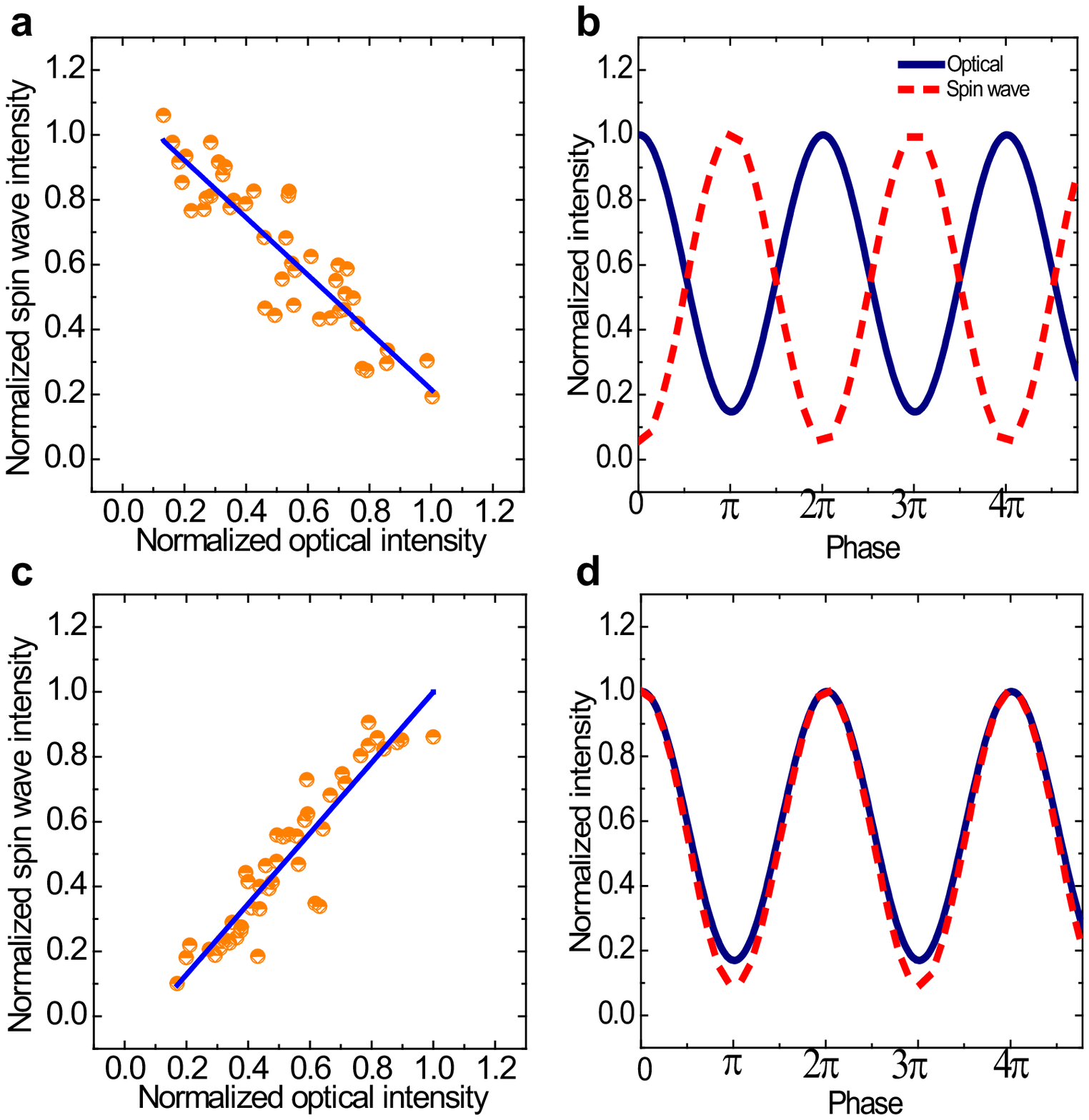} \caption{\textbf{The results of the pure Hermitian and non-Hermitian MPBS.}
The orange circles represents the smoothed experiments data, with blue
solid lines showing the fit correlation and anti-correlation. The
sine curve are theory fit experiment data corresponding interference
fringe. The navy solid lines represents the optical mode interference
fringes, and the red dash lines denote the atomic spin mode. (a)(b)
Pure coherent condition, while OD=40 and single photon detuning
is $\Delta/2\pi=60\,\mathrm{MHz}$. (c)(d) non-Hermitian MPBS condition,
with $\Delta=0$ and OD$=40$.}
\label{Fig3}
\end{figure}
\begin{figure*}
\includegraphics[width=0.9\textwidth]{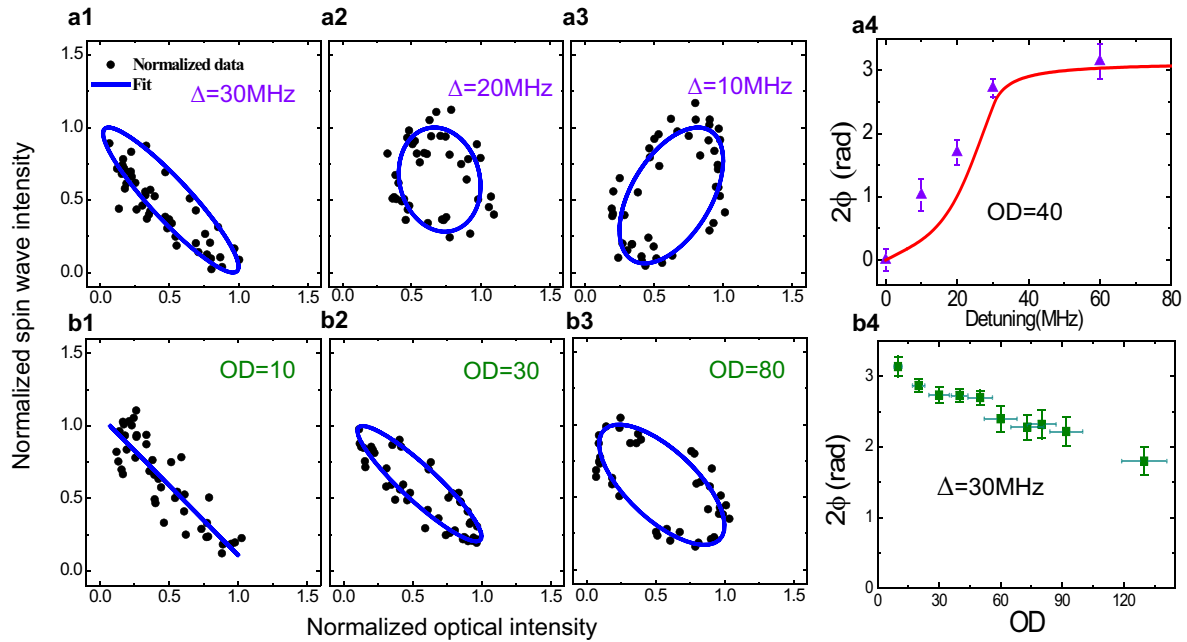} \caption{\textbf{The phase diagram for different experiment parameter of photon-magnon
interference.} The black dots represents the experiments data, and
the blue solid lines shows theory fit curve. The experiments data
were plotted through normalization processing. (a1)-(a4) OD$=40$
with different single photon detuning $\Delta$. (b1)-(b4) Single
photon detuning $\Delta/2\pi=30\,\mathrm{MHz}$ with different OD. }
\label{Fig4}
\end{figure*}
For a simplified model of our experiments based on single magnon mode,
the MPBS operation can be described the effective Hamiltonian $\left(\hbar=1\right)$
\cite{Hammerer}
\begin{align}
H_{\mathrm{eff}}= & \frac{g^{2}N}{\Delta-i\kappa_{13}}\hat{a}_{L}^{\dag}\hat{a}_{L}+\frac{\Omega_{c}^{2}}{\Delta-i\kappa_{13}}\hat{S}_{a}^{\dag}\hat{S}_{a}\nonumber \\
 & +\frac{g\sqrt{N}\Omega_{c}}{\Delta-i\kappa_{13}}\left(\hat{a}_{L}\hat{S}_{a}^{\dagger}-\hat{a}_{L}^{\dagger}\hat{S}_{a}\right)\label{eq:effective H}
\end{align}
under the approximations of $\left|\Delta\right|^{2}+\kappa_{13}^{2}\gg g,\Omega_{c}$
(See SM for more careful treatments). Here, $\Delta=\omega_{3}-\omega_{1}-\omega_{p}$
is the detuning, $\kappa_{13}$ represents the dephasing rate between
$\left|3\right\rangle $ and $\left|1\right\rangle $, $g$ is the
atom-photon coupling strength, $N$ is the number of atoms, and $\Omega_{c}$
is the Rabi frequency of the control field. In all experiments, the
signal and control light satisfies the two-photon resonance condition
, i.e., $\omega_{p}-\omega_{c}=\omega_{2}-\omega_{1}$. The last term
of Eq.$\,$(\ref{eq:effective H}) denotes the linear beam-splitter-type
of interaction \cite{Hammerer}, but with a complex coupling strength when
we adiabatically eliminate the excited state. The non-zero imaginary
part leads to the non-Hermitian \cite{El-Ganainy2018} photon-magnon
conversion in the cold atom ensemble. By considering a square signal
photon pulse with duration of $\tau_{p}$, the spatial-temporal input-output
function of the non-Hermitian MPBS reads
\begin{equation}
\left(\begin{array}{c}
S\left(\tau_{p}\right)\\
a\left(L\right)
\end{array}\right)=\left(\begin{array}{cc}
t & r\\
r^{\prime} & t^{\prime}
\end{array}\right)\left(\begin{array}{c}
S\left(0\right)\\
a\left(0\right)
\end{array}\right),\label{eq:T}
\end{equation}
with the transfer matrix elements $t=e^{-\frac{\zeta}{i\Delta/\kappa_{13}+1}}$,
$t^{\prime}=1-\frac{\eta}{\zeta}\left(1-t\right)$, $r=t-1$ and $r^{\prime}=\frac{\eta}{\zeta}r$.
Here, $\zeta=\Omega^{2}\tau_{p}/\kappa_{13}$ is a dimensionless number
quantify the beam-splitter interaction strength and $\eta$ is a parameter
depends on the OD.

From the transfer matrix, the parameter $\Delta/\kappa_{13}$ plays
an important role to control the coalescence of MPBS. With far-detuned
control, i.e., $\Delta/\kappa_{13}\gg1$, $T_{11}\approx1+i\zeta\kappa_{13}/\Delta$,
$T_{22}\approx1+i\eta\kappa_{13}/\Delta$, $T_{12}\approx i\zeta\kappa_{13}/\Delta$
and $T_{21}\approx i\eta\kappa_{13}/\Delta$, the MPBS therefore behaves
as a normal lossless BS with $\phi\approx\pi/2$. The phase difference
of the interference fringe is $\pi$, as shown in Fig.$\,$ \ref{Fig3}(a)
and (b). The output intensities at optical wave reaches the maximum
while the atomic spin port reaches the minimum value, and vice versa,
showing anti-correlation that conserves the number of excitations.
Oppositely, near-resonance excitation $\Delta/\kappa_{13}\rightarrow0$,
the MPBS is a lossy BS with off-diagonal terms of $T$ are negative
numbers. The outputs are correlated that tends to simultaneously reach
extreme values, and the phase difference of fringes is zero, as shown
in Fig. \ref{Fig3}$\,$(c) and (d). An implication of this result
is it is in the PT-symmetry broken regime of the non-Hermitian system
\cite{Zou2017,El-Ganainy2018}. There is only one eigenstate as a
superposition of photon and magnon, so the ratio $n_{a}/n_{s}$ is
a constant. For general $0<2\phi<\pi$, the $\left\{ n_{a},n_{s}\right\} $
show a ellipse trajectory.

As indicated by the Eq.$\,$(\ref{eq:T}), the phase difference of
interference fringes $2\phi$ can be adjusted by controlling the experiment
parameters of the MPBS, including the detuning $\Delta$, OD and coupling
power ($\Omega_{c}$). In Fig.$\,$ \ref{Fig4}, we performed more
detailed experimentally studied the phase of MPBS by reconstructing
the Lissajous curve. In Fig.$\,$ \ref{Fig4}(a1)-(a3), we decrease
single photon detuning $\Delta/2\pi$ from $30\,\mathrm{MHz}$ to
$0$ while maintaining OD$=40$. The experimental data are fitted
with blue solid ellipses, which turn into near-circular and change
its direction from anti-correlation to correlation with $\Delta/\kappa_{13}$
decreases. As summarized in Fig.$\,$ \ref{Fig4}(a4), with
theoretical lines calculated from full magnon-photon coupling model
{[}see Supplementary Materials{]}. Single mode approximation satisfactorily
describe the transition from Hermitian to Non-Hermitian. Fig. \ref{Fig4}(b1)-(b3)
represent the measured phase diagram for different ODs, with fixed
$\Delta/2\pi=30\,\mathrm{MHz}$, and they are summarized in Fig.$\,$ \ref{Fig4}(b4).
For increasing OD, the number of atoms $N\propto\mathrm{OD}$ increases,
where the single mode approximation breaks down {[}see Supplementary
Materials{]}. It is can be intuitively understand that a MPBS with
increased OD can be treated as a sequence of spatially cascaded single
magnon mode-based MPBS, thus the non-Hermitian induced phase would
increase with the OD. From Fig.$\,$ \ref{Fig4}(b4), it is clearly
demonstrated that the MPBS deviated from Hermitian, i.e. $2\phi$ decreases
from $\pi$, for growing $OD$. Thus, our MPBS can be easily reconfigurated
from a nearly-ideal Hermitian model to a non-Hermitian beam-splitter
by only changing the laser frequencies or the condition of MOT, would
be a convenient platform for future studies of non-Hermitian physics.

In conclusion, we demonstrate a tunable non-Hermitian MPBS in a three-level $\Lambda$
atomic system driven by a strong coupling field. Through changing the
optical depth and the single photon detuning ${\Delta}$, we can adjust its non-Hermiticity.
The non-Hermitian feature can be observed through interference between
optical field and collective atomic spin wave in an equivalent MZ-interferometer
setup. The phase difference of the interference fringes between the
optical mode and the atomic spin mode can be adjusted from $\pi$,
complementary to 0, correlated. A direct implication is the
bosons change their coalescence through this MPBS. This is caused by
the non-unitary transformation induced by incoherent photon-magnon interaction.
Finally, our work can extend to single-quantum level to realize Hong-Ou-Mandel interference
between a single photon and a single atomic spin mode. The non-Hermitian
atomic beam-splitter lay the foundation for verify P-T symmetry in
three-level EIT system, and how the quantum statistics of bosons and
fermions change in such exotic system.



\vspace{0.2in}

\vbox{}

\noindent \textbf{\large{}{}{}Methods}{\large\par}

A cigar-shaped laser cooled $^{85}$Rb atoms cloud is prepared in
a two-dimensional (2D) magneto-optical trap (MOT) with a length of
$1.5\,\mathrm{cm}$, the temperature of the MOT is about $100\,\mathrm{\mu K}$.
The cooling laser is red-detuned from the transition $5S_{1/2},F=3\leftrightarrow5P_{3/2},F=4$
by 18 MHz with power as 110 mw. The repump laser is on resonance with
the transition $5S_{1/2},F=2\leftrightarrow5P_{3/2},F=2$ which pumping
the atoms to the cooling cycle. The experiment periodically time is
5 ms which 4.5 ms as MOT preparation time and 0.5 ms as measurement
time window. We use three-level EIT system to perform the MPBS. The
probe and coupling laser, originated from the same diode laser, are
respectively frequency shifted using acousto-optics modulators. The
control laser (control, 795 nm) is blue detuned from the transition
$|2\rangle=5S_{1/2},F=3\leftrightarrow|3\rangle=5P_{1/2},F=3$ by
$\Delta$ and two-photon resonant with probe light (probe, 795 nm)
which is blue detunned from the $|1\rangle=5S_{1/2},F=2\leftrightarrow|3\rangle=5P_{1/2},F=3$.
They have same polarization $\sigma^{+}$ (or $\sigma^{-}$). The
coupling beams is collimated with a diameter of 1.6 mm which can make
the coupling beam recover whole cold atom cloud. The angle between
the coupling beam and the atom long axis direction is $2.5^{\circ}$.
A probe pulse (50ns) which created by the acousto-optic modulator
(AOM) through radio-frequrncy (RF) signal splitting into two well
separated pulses through a half wave plate, PBS, 20m polarization-maintained
single-mode fiber and fiber BS, then they are launched into MOT and
focused on the center of the cold atom cloud with diameter of 230$\mu$m,
the relative intensities of the two pulses can be modulated by half-wave
plate aimed to make power match of the optical and the spin wave.
The final optical signal collected by a photomultiplier tube (PMT).
The first coupling pulse width is 120ns, the second coupling pulse
width and the power are adjusted according to detuning and the OD
to modulation atomic beams plitter ratio, the third is extended to
600 ns with maximum intensity to make sure collective atomic mode
almost retrieved, the intensities of the three coupling pulse are
controlled by an EOM. The single photon detuning $\Delta$ controlled
by AOM and the OD through adjust the repump laser power.

The experimental procedure of the MPBS is shown in Fig$\,$\ref{Fig2}(c),
the two probe pulses (grey and pale green shadows) are generated by
a single $50\,\mathrm{ns}$ duration pulse ($300\,\mathrm{nW}$) that
split by an optical beam splitter, with a delay time of about $100\,\mathrm{ns}$
($20\,\mathrm{m}$ polarization-maintained single-mode fiber delay).
Then, a long control pulse with $600\,\mathrm{ns}$ (faint yellow
shodows in Fig$\,$\ref{Fig2}(c)) is applied to the atom ensemble
after the interference step, for both magnon readout and system initialization.
Figures~\ref{Fig2}(c)-(e) show the outputs of the MPBS for the case
of single photon detuning $30\,\mathrm{MHz}$ with an optical depth
(OD) of 30. In the following experiments, we adjusted the coupling
pulse intensities for the magnon preparation and interference steps
to match the amplitudes of output photon and magnon. Figure~\ref{Fig2}(d)
shows the results for single port input to MPBS (either photon or magnon
be vacuum state), which shows that the outputs at two ports matches.
Typical results for magnon-photon interference for different phase
$\phi$ are shown in Fig.~\ref{Fig2}(e), which demonstrate the change
of the outputs at different ports according to the interference. Here,
since our real experimental time (about $1\,\mathrm{\mu s}$) is very
short compared with the cycling time ($5\,\mathrm{ms}$) and the phase
uncertainty of the fiber is negligible {[} according to \cite{Lucamarini2018},
phase drift during single experiment cycle$<18\,\mathrm{rad/ms/200km}\times20\,\mathrm{m}\times1\,\mathrm{\mu s}\sim1.8\times10^{-6}\,\mathrm{rad}${]},
the relative phase of the probe pulses is not controlled manually
but fluctuate randomly. Therefore, we record the output intensities
of photon and magnon in the phase diagram, where the intensity are
obtained by integrating the areas of the output pulses (grey region
and the rosiness region) in the Fig.~\ref{Fig2}(e).

\vbox{}

\noindent \textbf{Acknowledgments}\\
This work is supported by the National Key Research and Development
Program of China under Grant number 2016YFA0302001, and the National
Natural Science Foundation of China through Grant No. 11674100, 11654005,
11234003, the Natural Science Foundation of Shanghai No. 16ZR1448200,
and Shanghai Rising-Star Program 17QA1401300.







\clearpage{}

\newpage{}

\newpage{}

\onecolumngrid
\global\long\def\thefigure{S\arabic{figure}}%
 \setcounter{figure}{0}
\global\long\def\thepage{S\arabic{page}}%
 \setcounter{page}{1}
\global\long\def\theequation{S.\arabic{equation}}%
 \setcounter{equation}{0} 
\setcounter{section}{0}
\end{document}